# Speech: A Challenge to Digital Signal Processing Technology for Human-to-Computer Interaction

Urmila Shrawankar
Dept. of Information Technology
Govt. Polytechnic, Nagpur Institute
Sadar, Nagpur – 440001 (INDIA)
urmilas@rediffmail.com
Cell : (+91) 9422803996

Anjali Mahajan
Dept. of Computer Sci & Engg.
G H Raisoni College of Engg.,
Hingna, Nagpur 440016 – (INDIA)
armahajan@rediffmail.com
Phone: (0712)-2241509 ®

*Abstract* - **This software project based paper is for a vision of the near future in which computer interaction is characterised by natural face-to-face conversations with lifelike characters that speak, emote, and gesture. The first step is speech. The dream of a true virtual reality, a complete human-computer interaction system will not come true unless we try to give some perception to machine and make it perceive the outside world as humans communicate with each other. This software project is under development for "listening and replying machine (Computer) through speech".**
**The Speech interface is developed to convert speech input into some parametric form (Speech-to-Text) for further processing and the results, text output to speech synthesis (Text-to-Speech)**

## I. INTRODUCTION

When we think of user interfaces, the very first question arises in the mind is that why do we need an interface to interact with a machine (Computer)? The answer is simple, human-to-computer interaction is not simple as human-to-human interaction. Human-to-human interaction mainly based on speech, emotion and gesture, where as Human-to-machine interaction based on either Text User Interface (TUI) or Graphical User Interface (GUI).

If we provide an artificial intelligence to train a machine in such a way, so that machine will interact using speech signals. This paper will focus on developing software based user interface to accept speech input through microphone and gives speech output through speakers connected to computer.

My try is to develop a speaker independent and text dependent model i.e. after completing the training from variety of samples computer will able to understand any type of voice such as male, female, children of any age group and for specific text.

Speech recognition system helps user, who are unable to use the traditional Input and Output (I/O) devices. Since four decades, human beings have been dreaming of an "intelligent machine" which can master the natural speech. In its simplest form, this machine should consist of two subsystems, namely automatic speech recognition (ASR) and speech understanding (SU). The goal of ASR is to transcribe natural speech while SU is to understand the meaning of the transcription. Recognizing and understanding a spoken sentence is obviously a knowledge-intensive process, which must take into account all variable information about the speech communication process, from acoustics to semantics and pragmatics. The model of a Speech recognition system is as below. (**Fig. Speech Model**)

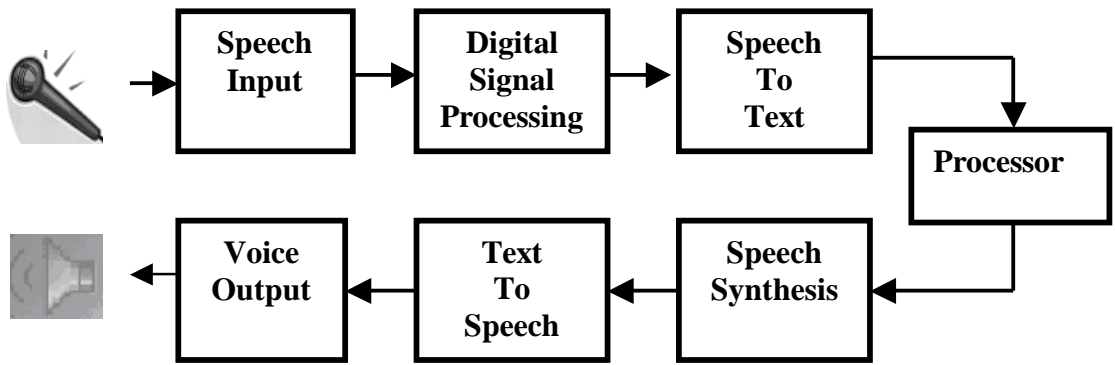

**Fig. SpeechModel : The Speech Interface Model**

## II. ROLE OF DIGITAL SIGNAL PROCESSING IN SPEECH RECOGNITION

Signal processing is the process of extracting relevant information from the speech signal in an efficient, robust manner.

A speech recognition system comprises a collection of algorithms drawn from a wide variety of disciplines, including statistical pattern recognition, communication theory, signal processing, combinational mathematics, and linguistics, among others. Although each of these areas is relied on to varying degrees in different recognizers, perhaps the greatest common denominator of all recognition systems is the signal processing front end, which converts the speech waveform to some type of parametric representation for further analysis and processing.

### A. Speech Signal Processing

Speech recognition can be defined as the process of converting an acoustic signal, captured by a microphone or a telephone, to a set of words.

After text-to-speech (TTS) and interactive voice response (IVR) systems, automatic speech recognition (ASR) is one of the fastest developing fields in the framework of speech science and engineering. As the new generation of computing technology, it comes as the next major innovation in man-machine interaction. Speech recognition systems can recognize thousands of words. The evolution of ASR has a lot of applications in many aspects of our daily life, for example, telephone applications, applications for the physically handicapped and illiterates and many others in the field of computer science. Speech recognition is considered as an input as well as an output during the Human Computer Interaction (HCI) design. HCI involves the design implementation and evaluation of interactive systems in the context of the users' task and work

### B. Speech Recognition Systems

Speech Recognition is a technology, which allows control of machines by voice in the form of isolated or connected word sequences. It involves the recognition and understanding of spoken language by machine.

Speech Recognition is based on a pattern recognition technology. The objective is to take an input pattern, the speech signal and classify it as a sequence of stored patterns that have precisely been defined. These stored patterns may be made of units, which we call *phonemes.*

If speech patterns were invariant and unchanging, there would be no problem; simply compare sequences of features with the stored patterns, and find exact matches when they occur. But the fundamental difficulty of speech recognition is that the speech signal is highly variable due to different speakers, different speaking rates, different contents and different acoustic conditions. The task is to determine which of the variations in the speech are relevant to speech recognition and which variations are not relevant.

## III. FEATURE EXTRACTIONS AND FEATURE MATCHING

Feature extraction is the process that extracts a small amount of data from the voice that can later be used to represent each word. Feature matching involves the actual procedure to identify the new word by comparing extracted features from his/her voice input with the ones from a set of known words.

All speech recognition systems have to serve two distinguishes phases. The first one is the enrollment sessions or training phase while the other is the testing phase.

### A. Speech Feature Extraction

The purpose of this module is to convert the speech waveform to some type of parametric representation for further analysis and processing. This is often referred as the *signal-processing front end.*

A wide range of possibilities exist for parametrically representing the speech signal and the speech recognition task, such as Mel-Frequency Cepstrum Coefficients (MFCC), Linear Prediction Coding (LPC), Filter-bank Spectrum analysis model, Vector Quantisation and others. The LPC model is implemented in this project.

### B. Linear Predictive Coding (LPC) Model

Linear Predictive Coding (**LPC**) is one of the most powerful speech analysis techniques and a useful method for encoding quality speech at a low bit rate. It provides accurate estimates of speech parameters and efficient for computations.

LPC system is used to determine the formants from the speech signal. The basic solution is a difference equation, which expresses each sample of the signal as a linear combination of previous samples. Such an equation is called a linear predictor that is why this is called Linear Predictive Coding.

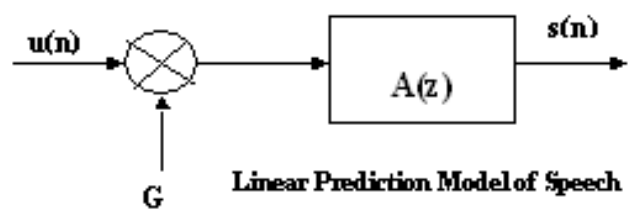

Linear Prediction Model of Speech

The basic idea behind the LPC model is that a given speech sample at time n, s(n), can be approximated as a linear combination of the past p speech samples.

After completing steps as shown in the **fig. LPC** we get parameters from the speech signal, further these are used for training purpose.

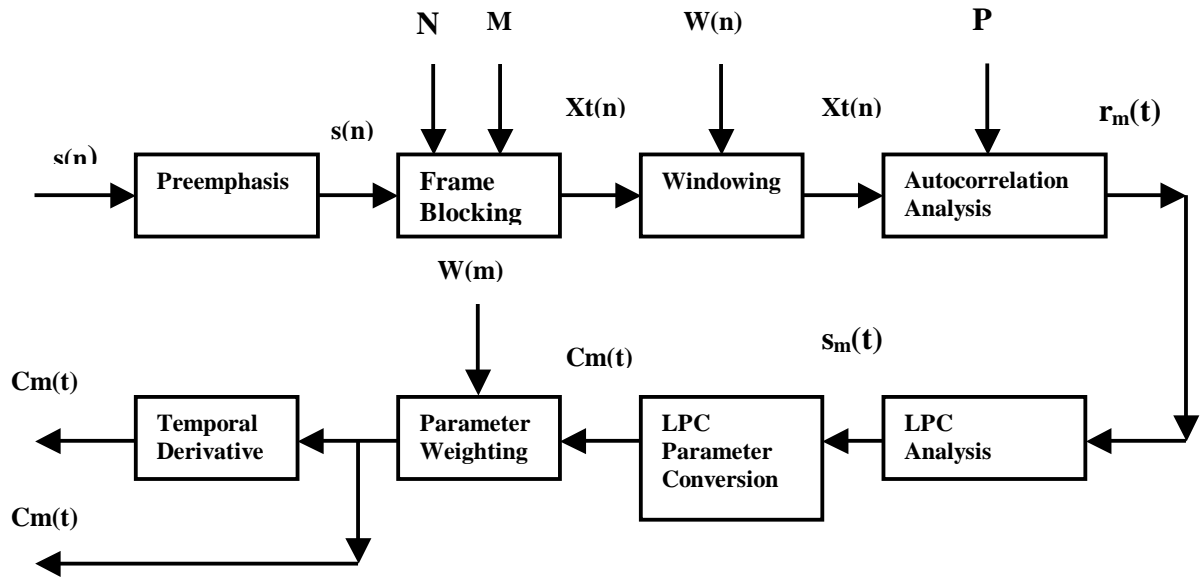

Fig. LPC : Block diagram of the LPC processor

## IV. TRAINING and RECOGNITION

In the training phase a data file is created and the samples that are recorded from different users are stored. These samples are further matched and use to recognise the word.

### A. Training with Artificial Neural Networks

Neural networks are often used as a powerful discriminating classifier for tasks in automatic speech recognition. They have several advantages over parametric classifiers. However, there are disadvantages in terms of amount of training data required, and length of training time. Some neural network architectures are:

- Feedforward Perceptrons Trained With BackPropagation
- Radial Basis Function (RBF) Networks
- Learning Vector Quantization (LVQ) Networks

### B. Training with Hidden Markov Model

In the context of statistical methods for speech recognition, Hidden Markov Models (HMM) have become a well known and widely used statistical approach to characterising the spectral properties of frames of speech. Hidden Markov Model is a *doubly stochastic process* in which the observed data are viewed as the result of having passed the true (hidden) process through a function that produces the second process (observed). The hidden process consists of a collection of states (which are presumed abstractly to correspond to states of the speech production process) connected by transitions. (**Refer fig HMM**) Each transition is described by two sets of probabilities:
• A *transition probability*, which provides the probability of making a transition from one state to another.
• An *output probability* density function, which defines the conditional probability of observing a set of speech features when a particular transition takes place.
The goal of the decoding (or recognition) process in HMMs is to determine a sequence of (hidden) states (or transitions) that the observed signal has gone through. The second goal is to define the likelihood of observing that particular event given a state determined in the first process. The Isolated Word Recognition model is shown in **fig. RU**.

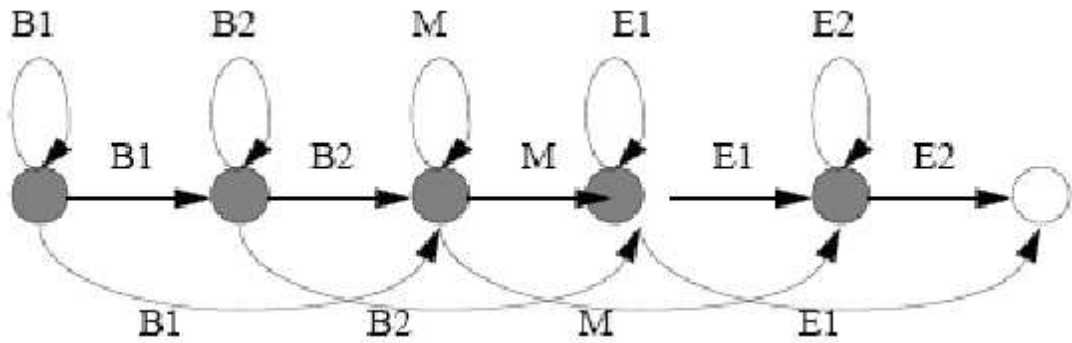

**Fig. HMM : The topology of the phonetic HMM**

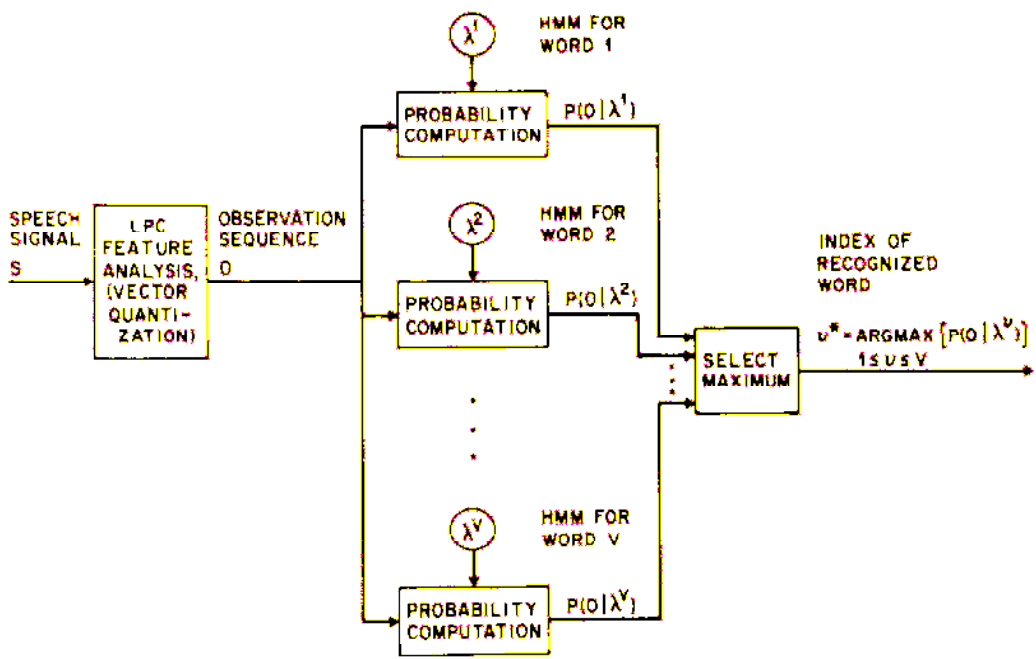

**Fig. RU: Recognition Unit: HMM System For Isolated Word Recognition**

The training procedure involves optimizing HMM parameters given an ensemble of training data.
Following algorithms are implemented for training and recognizing isolated words.
- Forward- Backward Algorithm
- Viterbi Algorithm.
- Baum-Welch Algorithm

An iterative procedure, the Baum-Welch or forward-backward algorithm, is employed to estimate transition probabilities. The Viterbi algorithm is used as a fast-match algorithm. The decoder is designed to exploit all available acoustic and linguistic knowledge in several search phases.
Using HMMs by giving a set of performance results on the task of recognizing isolated digits in the speaker independent manner.

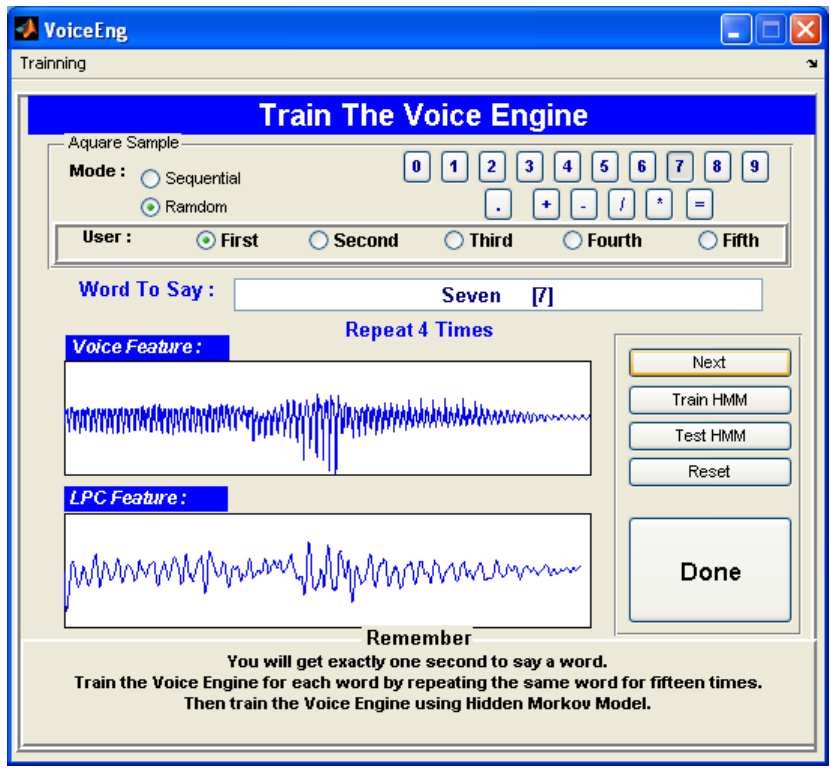

Fig. VoiceEng :The Voice Engine Software Front-End

## V. SPEECH SYNTHESIS (TEXT-TO-SPEECH)

Speech synthesis is the reverse process to the recognition. The advances in this area improve the computers' usability for visually impaired people

*Text-to-phoneme conversion:* Once the synthesis processor has determined the set of words to be spoken, it must derive pronunciations for each word. Word pronunciations may be conveniently described as sequences of phonemes, which are units of sound in a language that serve to distinguish one word from another.

## VI. THE SOFTWARE

*VOICE ENGINE*
Software is developed for training and testing Speech Interface. As a sample application, the machine is trained for numbers (0-9), and some mathematical operators.
The input numbers and operators are provided to machine through microphone in the form of wav files. Features are extracted from these speech signals and passed to parametric forms for further processing. Extracted parameters are sent to training unit.
Total 25 samples are collected for training from 5 different age-group users (Males and Females), five attempts from each user for every word. The generated results are outputted through speaker.

## VI. ADVERSE CONDITIONS IN SPEECH RECOGNITION

While developing this project, it is observed that some adverse conditions degrade the performance of the Speech Recognition system.

### A. Noise

If we use the noise free environment to train and test we get, about 80% accuracy. But if the room is noisy either in training phase or in testing phase accuracy is reduces to around 60%

### B. Distortion

To implement this project we do not require any special hardware other than the computer machine, a Microphone , speakers or a headphone with microphone. If these attachments are not installed and configure properly we get distorted input signals, which reduces the accuracy.

### C. (Human) Articulation Effects

Many factors affect the manner of speaking of each individual, like the distance of microphone from the user and its position, also, speech added with psychological effect while providing input, these factors effects the accuracy.

## VIII. CONCLUSIONS:

A technology without social aspect is useless. Now-a-days, computer became a part of day-to-day working. After getting the graphical user interface (GUI) it is very easy to interact with the computer. But still it is difficult to interact for the physically challenged people especially for *Amelia* (Absence of limb-handless) person and *Blinds* or *old age people*.

The speech is an interface for Human-Computer interaction, so that they can interact with computer without keyboard or mouse.

This project software is developed for partially fulfillment of M.Tech. (Computer Science and Engineering) degree.

The software is developed using VoiceBox and H2M toolbox of MatLab 7.1 and based on Artificial intelligence Speech Recognition approach.

No other computer hardware is required except microphone and speakers.

This software gives 80% accuracy with clean environment and about 60% with noisy environment.